\documentstyle[11pt,epsfig,epsf]{article}
\def\slash#1{\setbox0=\hbox{$#1$}#1\hskip-\wd0\dimen0=5pt\advance
\dimen0 by-\ht0\advance\dimen0 by\dp0\lower0.5\dimen0\hbox
to\wd0{\hss\sl/\/\hss}}

\newcommand{\sm}{s_-}
\newcommand{\ssp}{s_+}
\newcommand{\dm}{\delta m}
\newcommand{\dmb}{m_{B^*}-m_B}
\newcommand{\dmd}{m_{D^*}-m_D}
\newcommand{\eps}{\Gamma_{D^{*}}}
\newcommand{\om}{\omega}
\newcommand{\fhat}{\hat F}
\newcommand{\fhatpiu}{\hat F^+}
\newcommand{\vpw}{v \cdot v_-}
\newcommand{\vvp}{v \cdot v_+}
\newcommand{\vw}{v_+ \cdot v_-}
\newcommand{\qv}{q \cdot v_+}
\newcommand{\qw}{q \cdot v_-}
\newcommand{\qvp}{q\cdot v}

\newcommand{\be}{\begin{equation}}
\newcommand{\ee}{\end{equation}}
\newcommand{\bea}{\begin{eqnarray}}

\newcommand{\eea}{\end{eqnarray}}
\newcommand{\nn}{\nonumber }
\newcommand{\dd}{\displaystyle }

\begin{document}

\title{\hfill 
$\mbox{\small{\begin{tabular}{r}
${\textstyle Bari-TH/98-316}$
\end{tabular}}}$ \\[1truecm]
{\bf Analysis of the Three-Body ${\mathbf B \to D^+ D^- \pi^0}$ Decay }}
\author{P. Colangelo$^a$, F. De Fazio$^a$, G. Nardulli$^{a,b}$, 
N. Paver$^c$ and Riazuddin$^d$} 
\maketitle
%
\begin{it}
\noindent
$^a$ Istituto Nazionale di Fisica Nucleare, Sezione di Bari, Italy\\
$^b$ Dipartimento  di Fisica dell'Universit\`a di Bari, Italy\\
$^c$ Dipartimento  di Fisica Teorica dell'Universit\`a di Trieste and
INFN, Sezione di Trieste, Italy\\
$^d$ Department of Physics, King Fahd University of Petroleum and Minerals, 
Dhahran, Saudi Arabia
\end{it}
\begin{abstract}
The decay process $B\to D^+ D^- \pi^0$ is an interesting channel
for the investigation of CP violating effects in the $b-$ sector.
We write down a decay amplitude constrained by a low-energy theorem,  which 
also includes the contribution of 
resonant $S-$ and $P-$wave beauty and charmed 
mesons, and we determine the relevant matrix elements in the infinite
heavy quark mass limit, assuming the factorization ansatz. 
We estimate the rate of the decay:
${\cal B}(B \to D^+ D^- \pi^0)\simeq 1 \times 10^{-3}$.
We also analyze the time-independent and 
time-dependent  differential decay distributions,  concluding that 
a signal for this  process should be observed at the B-factories.
Finally, we give an estimate of the decay rate of the Cabibbo-favoured 
process $B\to D^+ D^- K_S$. 
\end{abstract}
\newpage
\section{Introduction}\label{s:intr}

Multibody hadronic $B$ decays represent a large fraction of the inclusive 
nonleptonic rate \cite{PDG}, and therefore it is worth analyzing their 
phenomenological aspects,  since they constitute accessible channels for the 
experimental investigations \cite{review}. In particular, some three-body 
neutral $B$ hadronic decays deserve further attention, 
from both the theoretical and the 
experimental sides, since they have been recognized as an important source of 
information concerning CP violation in the beauty sector. 
This is the case of the 
decays ${\overline B^0}\to \pi^+ \pi^- \pi^0$ and
$B^0\to \pi^+ \pi^- \pi^0$, which provide information,
together with the neutral $B$  decays into a pion pair, 
on the CKM angle $\alpha$
\cite{review, quinn}. This is also the case of the three-body decays
${\overline B^0}\to D^+ D^- \pi^0$ and $B^0 \to D^+ D^- \pi^0$, 
which have been identified as 
interesting channels to investigate the angle $\beta$ \cite{charles}, in 
particular as far as the discrete ambiguity 
$\beta \to {\pi \over 2} - \beta$ of the CP asymmetry 
in $B \to J/\psi K_S$ is concerned. The removal of such ambiguities and, in 
general, the identification of possible constraints on the CKM angles are of 
prime interest, mainly in view of testing the Standard Model and probing the 
effects of new physics scenarios. \cite{buras,discr}

The theoretical calculation of multibody hadronic $B$ decays presents 
uncertainties, mainly related to the long-distance QCD effects involved in 
these 
processes. Simplifying assumptions are usually adopted, such as, for example, 
the hypothesis of dominance of intermediate hadronic resonances in the 
relevant 
amplitudes. In the case of pions in the final state, however, low-energy 
theorems can be employed to reduce the decay amplitude in the soft-pion 
limit $q_\pi \to 0$; this allows, for example, to relate a three-body decay 
amplitude to a corresponding two-body one. If a narrow phase space is available
around the  point $q_\pi \to 0$, an extrapolation can be 
done to estimate the multibody process. This program cannot be pursued for the 
$B \to 3 \pi$ decays, where high momentum pions are allowed 
in the final state. 
The situation presents less difficulties in the case of the decay 
$B \to D^+ D^- \pi^0$, where a quite narrow phase space 
is available for the pion;
therefore, the amplitude having the right behaviour for $q_\pi \to 0$
can be extrapolated, including the contribution of
intermediate resonant states, to the full phase space.

This is the aim of the present work.  We shall write down an 
amplitude for $B \to D^+ D^- \pi^0$ and for the $SU(3)$-related process
$B \to D^+ D^- K_S$  having the soft pion limit
required by current algebra and PCAC, and
including a set of intermediate hadronic states.
In this way, the amplitude can be reduced to a set of two-body hadronic matrix 
elements,
which we shall evaluate by the factorization ansatz; the description will be 
simplified by observing that, in the infinite heavy quark ($b,c$) mass limit, 
the hadronic matrix elements involved in the calculation are related to a few
universal (mass-independent) parameters.
An interesting observation will 
be that the full amplitude can be derived from an
effective Lagrangian, obeying chiral symmetry in the light meson sector and 
heavy quark symmetry in the heavy quark sector. The unknown parameters are the
Isgur-Wise semileptonic form factors, the heavy meson leptonic constants and 
the effective couplings describing the QCD interactions of the heavy mesons 
with pions.

The plan of the paper is as follows.
In Section \ref{s:kin} we briefly review the kinematics of the 
$B \to D^+ D^- \pi^0$ decay, and the relevance of this channel 
in the perspective of CP measurements. 
In Section \ref{s:low} 
we derive the low-energy theorem for the nonleptonic 
$B \to D^+ D^- \pi^0$ amplitude, 
together with the contribution of intermediate resonant states, and provide
an  evaluation of such an amplitude by using the factorization ansatz.
In Section \ref{s:efflag} we discuss a derivation based on 
an effective heavy meson chiral Lagrangian.
The numerical analysis is reported in Section \ref{s:res}, and a short 
discussion of the Cabibbo-favoured  $B \to D^+ D^- K_S$ decay 
concludes the presentation.
\section{Kinematics and $\beta$ dependence}\label{s:kin}
We consider the process:
\be
{\overline B^0}(p) \to D^+(p_+)~D^-(p_-)~\pi^0 (q)  \label{process}
\ee
and the analogous one for the $B^0$ meson.
Neglecting penguin contributions, these decays are  governed by the weak
Hamiltonian
\be
H_W= {G_F \over \sqrt 2} V_{cb} V^*_{cd} \big(c_1 + {c_2 \over N_c}\big)
\bar b \gamma^\mu (1- \gamma_5) c \; \bar c \gamma_\mu (1-\gamma_5) d \;\;+ 
h.c.\;\;\;;
\label{eq:hamilt}
\ee
$G_F$ is the Fermi constant, $V_{ij}$ are  CKM  matrix elements
and $c_{1,2}$ are short-distance coefficients, with $N_c$ the number of
colours.
 
The neglect of gluon and electroweak penguin operators, that in principle 
contribute to  process (\ref{process}), cannot be
justified {\it a priori}. However, considering that the corresponding
short-distance coefficients are rather small, and that  process
(\ref{process}) is colour-allowed, we may expect the dominance of the
"tree-diagram" operator in Eq.(\ref{eq:hamilt}) 
to be a good approximation in this
case. Qualitative estimates based on the factorization ansatz suggest
for the two-body $b \to c \bar c d$ decays that
corrections from the penguin contributions could be of the order of few
percents \cite{deshpande}. We recall that, due to the
common $\Delta I= 1/2$ character of both tree and penguin operators
for the transition $b \to c \bar c d$, the corresponding amplitudes cannot 
be separately determined by the isospin analysis \cite{sanda}.

Following the notations of {Ref.~\cite{charles}}, we define the Dalitz plot
variables of the decay (\ref{process}):
\begin{eqnarray}
s_+ &= & (p_+ +q)^2 \nn \\
s_- &= & (p_- +q)^2  \\
s_0 &= & (p_+ +p_-)^2=m_B^2+2m_D^2+m_\pi^2-s_+ -s_-
 ~~~~.\nn
\end{eqnarray}
In terms of the heavy meson four-velocities 
\be p^{\mu}=m_B v^{\mu}~~~~,~~~~
p_+^\mu=m_D v_+^\mu~~~~,~~~~
p_-^\mu=m_D v_-^\mu~~~~,
\ee
we also introduce a set of invariant variables, suitable for the
application of the heavy quark effective theory (HQET)  \cite{HQET}
to our problem: 
\bea
q\cdot v_+ &=&\frac{\ssp-m_D^2-m_\pi^2}{2 m_D}      \nn\\
q \cdot v_- &=&\frac{\sm-m_D^2-m_\pi^2}{2 m_D}      \nn\\
q \cdot v &=&\frac{\ssp+\sm-2 m_D^2}{2 m_B}            \\
v_+ \cdot v_- &=&\frac{m_B^2 + m_\pi^2-\ssp-\sm}{2 m_D^2}   \nn\\
v \cdot v_+ &=&\frac{m_B^2+m_D^2-\sm}{2 m_D m_B}  \nn\\
v \cdot v_- &=&\frac{m_B^2+m_D^2-\ssp}{2 m_D m_B} \;\;\;\;. \nn
\label{eq:invariants}
\eea
In the plane $(s_-,s_+)$ the allowed kinematical region is bounded by the curves
\begin{eqnarray}
s_+^{min} &=& \left(E_2+E_3\right )^2-\left(\sqrt{E_2^2-m_\pi^2}+ 
\sqrt{E_3^2-m_D^2}\right)^2 \label{dalp}\\
s_+^{max} &=& \left(E_2+E_3\right)^2-\left(\sqrt{E_2^2-m_\pi^2} - 
\sqrt{E_3^2-m_D^2}\right)^2 \label{dalm}
\end{eqnarray}
where $\dd {E_2=\frac{s_- - m_D^2+m_\pi^2}{2\sqrt{s_-}}}$ and 
$\dd {E_3=\frac{ -s_- - m_D^2+m_B^2}{2\sqrt{s_-}}}$,
with $(m_D+m_\pi)^2 \le s_- \le (m_B-m_D)^2$.
The kinematical region  is symmetric under the exchange 
$s_- \leftrightarrow s_+$; CP
eigenstates correspond to the line $s_+=s_-$.

The role of three-body decays in accessing the weak angle $\beta$
has been discussed in Ref. \cite{charles}, and we repeat here the basic
points and the notations of that  
analysis. Since no weak phases appear in the effective Hamiltonian
(\ref{eq:hamilt}) in the Wolfenstein parametrization, the only
relevant weak phase in the $B^0$ (and $\overline B^0$) decays of the kind  
(\ref{process}) is the 
phase $\beta$ of the  mixing $B^0-{\overline B}^0$. 
Denoting, as in Ref. \cite{charles}, by
${\cal A}(s_+,s_-)$ and   
${\overline {\cal A}} (s_+,s_-)$ the amplitude for the decay
into $D^+ D^- \pi^0$
of the $B^0$ and 
$\overline B^0$, respectively,  
the time-dependent decay probability of a state 
identified  as a $ B^0$ at $t=0$ is given by:
\bea
| A (B^0(t) \to D^+ D^- \pi^0)|^2 &=&
{e^{- \Gamma t} \over 2}
\Big[ G_0(s_+,s_-) + G_c(s_+,s_-) \cos (\Delta m t) \nn\\
&-& G_s(s_+,s_-) \sin (\Delta m t) \Big]
\;, \label{eq:at}
\eea 
where
\bea
G_0(s_+,s_-)&=&|{\cal A}(s_+,s_-)|^2+|{\overline {\cal A}}(s_+,s_-)|^2 \nn\\
G_c (s_+,s_-)&=&|{\cal A}(s_+,s_-)|^2-|{\overline {\cal A}}(s_+,s_-)|^2 \nn\\
G_s (s_+,s_-)&=&
-2 \sin(2 \beta) Re ({\cal A}^*(s_+,s_-) {\overline {\cal A}}(s_+,s_-))
+2 \cos(2 \beta) Im ({\cal A}^*(s_+,s_-) {\overline {\cal A}}(s_+,s_-)) \nn\\
&=& -2 \sin(2 \beta) Re \tilde G_s (s_+,s_-) 
+2 \cos(2 \beta) Im \tilde G_s (s_+,s_-) \;\;\;.
\label{eq:gis}
\eea
For the analogous decay
of the ${\overline B}^0$ one has 
\bea
| A ({\overline B}^0(t) \to D^+ D^- \pi^0)|^2 &=&
{e^{- \Gamma t} \over 2}
\Big[ G_0(s_-,s_+) - G_c(s_-,s_+) \cos (\Delta m t) \nn\\
&+& G_s(s_-,s_+) \sin (\Delta m t) \Big]
\;. \label{eq:at1}
\eea 
Assuming that no direct CP violation occurs, consistently with the neglect
of penguin operators, the condition
${\overline {\cal A}}(s_+,s_-)={\cal A}(s_-,s_+)$ is verified;
in this case, the
time-independent term $G_0(s_+,s_-)$ in Eqs.(\ref{eq:at})-(\ref{eq:at1}) is
symmetric in $s_+ \leftrightarrow s_-$, while the coefficient
$G_c(s_+,s_-)$ of the $\cos (\Delta m t)$ term is antisymmetric. 
In principle, such
contributions to the decay rate can be directely tested by symmetric or, 
respectively,
antisymmetric integration over $s_+,s_-$ of the time-dependent Dalitz plot
distribution of events. 

As far as the interference term $\sin (\Delta m t)$
in  (\ref{eq:at})-(\ref{eq:at1}) is concerned,  $G_s(s_+,s_-)$ would be
symmetric under  $s_+ \leftrightarrow s_-$ in the case of real 
$A(s_+,s_-)$, and only the "indirect" CP violating part proportional to
$\sin(2 \beta)$ would survive. In principle, also
 the effect of this term can  be
disentangled from the other ones by
symmetric integration in $s_+$ and $s_-$ of the experimental 
time-dependent Dalitz plot.
In the general case, however, the amplitude
$A(s_+,s_-)$ will have a
non-trivial  CP-conserving phase $\delta(s_+,s_-)$, of strong interaction
origin, such that both the CP-violating  $\sin(2 \beta)$ term and the
CP-conserving $\cos(2 \beta)$ term will contribute to $G_s(s_+,s_-)$. In
particular, the role of the latter term was emphasized in Ref.
\cite{charles}, following the discussion of the decay process
$B \to 3 \pi$ of Ref.\cite{quinn}, as a possible resolution of the discrete
ambiguity $\beta \to {\pi\over 2} - \beta$ implicit in the experimental 
determination of 
$\sin(2\beta)$ from, e.g., the time-dependent CP asymmetry of the process 
$B\to J/\psi K_S$. One can easily see that, in the hypothesis of
no direct CP-violation, 
  the $\sin(2 \beta)$ component of $G_s(s_+,s_-)$ should be
symmetric under $s_+ \leftrightarrow s_-$ as being proportional to 
$\cos(\delta(s_+,s_-)- \delta(s_-,s_+))$, whereas the  $\cos(2 \beta)$  term
should be antisymmetric as being proportional to 
$\sin (\delta(s_+,s_-)- \delta(s_-,s_+))$. Thus, 
under such assumption they could
be disentangled by symmetric and, respectively, antisymmetric integration
of the $\sin (\Delta m t)$ component of the Dalitz plot distribution.

According to the above considerations,
Eqs.(\ref{eq:at}) and (\ref{eq:at1}) imply 
that a time-dependent analysis of the neutral $B$ 
decay to $D^+ D^- \pi^0$ gives access to $\cos(2 \beta)$ if the product
${\cal A}^* {\overline {\cal A}}$ has a non-vanishing imaginary part.
Clearly, the required CP-conserving strong phase between 
${\cal A}(s_+,s_-)$ and ${\bar{\cal A}}(s_+,s_-)$ must have 
a non-trivial dependence on $s_+$ and $s_-$. 
Following Ref. \cite{charles}, we 
assume the variation of such strong interaction phase over the 
Dalitz plot to be entirely determined by a set of excited $B^*$ and $D^*$ 
resonance contributions, parameterized by Breit-Wigner 
poles in the relevant channels. In addition, however, considering the rather
low energies (on the heavy quark mass scale)
allowed to the pion in the considered decay,  
we constrain such polar expression for the decay amplitudes to obey the 
low energy theorem resulting from chiral symmetry.
In the appropriate limit, $q \to 0$, the amplitude 
reduces to the continuum ``contact'' term determined by the 
general (and model independent) current algebra procedure. 
Unavoidably, the assumed resonance behaviour in $s_+$ and $s_-$, as well
as the factorization approximation for the relevant two-body matrix
elements, introduce some amount of  
model dependence that is difficult to
reliably assess on purely theoretical grounds. On the other hand,  the
experimental study 
of the Dalitz plot distribution of events 
should allow to test the phenomenological validity of the model, and in 
particular to evidence non-resonant contributions to the strong phase 
variation if they turned out to be large.\footnote{The 
sources of systematic uncertainties implicit in the assumed resonance 
parameterization of the strong phase behavior have been discussed in detail 
for the three-body $B\to\rho\pi\to 3\pi$ decay in \cite{quinn}.}
In any case, we shall estimate the theoretical uncertainties of our approach
by considering two different extrapolations for the Breit-Wigner poles.  
They will be discussed in the next Sections.
\section{Low energy theorem and polar contributions}\label{s:low}
In order to derive a low energy theorem for the
amplitude (\ref{process}) we consider the Ward identity
\cite{riaz}
\begin{eqnarray}
{\overline {\cal A}}
&=& <D^+(p_+) D^-(p_-) \pi^0(q)|H_W|{\bar B}^0(p)> \nonumber \\
&=& -i {\sqrt{2} \over f_\pi} < D^+ (p_+) D^-(p_-) | [F_5^3, H_W]| 
{\bar B}^0(p)> +lim_{q \to 0} \Big[-i {\sqrt{2} \over f_\pi} 
q^\lambda M_\lambda - {\overline {\cal A}}_B\Big] \nonumber \\
&+& {\overline {\cal A}}_B +{\overline {\cal A}}_R \; . \label{ward}
\end{eqnarray}
\noindent In Eq. (\ref{ward}), $M_\lambda$ is
\begin{equation}
M_\lambda=i\; \int d^4x e^{i q \cdot x}< D^+ (p_+) D^-(p_-)| \Theta (x_0)
[A_\lambda(x),H_W(0)]|{\bar B}^0(p)> \;\;\; ,  \label{mlambda}
\end{equation}
with $A_\lambda= 
(\bar u \gamma_\lambda \gamma_5 u - 
\bar d \gamma_\lambda \gamma_5 d)/2$ and $F_5^3$ the corresponding axial 
charge; $H_W$ is the weak Hamiltonian in Eq. (\ref{eq:hamilt})
and $f_\pi=132$ MeV.
The polar contributions ${\overline {\cal A}}_B$ and 
${\overline {\cal A}}_R$ in Eq. (\ref{ward}) are
depicted in fig.1. 
%
\begin{figure}[h]
\begin{center}
\input feynman.tex
\begin{picture}(34000,10000)
\THICKLINES
\drawline\fermion[\E\REG](0,0)[4500]
\drawarrow[\LDIR\ATTIP](\pmidx,\pmidy)
\global\advance\pmidy by -1500
\put(\pmidx,\pmidy){$B$}
\put(4500,0){\circle*{800}}
\drawline\scalar[\NE\REG](\pbackx,\pbacky)[2]
\drawarrow[\LDIR\ATTIP](\pmidx,\pmidy)
\global\advance\pmidx by 2000
\global\advance\pmidy by 1600
\put(\pmidx,\pmidy){$\pi$}
\drawline\fermion[\E\REG](\pfrontx,\pfronty)[4500]
\drawarrow[\LDIR\ATTIP](\pmidx,\pmidy)
\global\advance\pmidx by -1000
\global\advance\pmidy by -1500
\put(\pmidx,\pmidy){$P_b$}
\put(\pbackx,\pbacky){\rule[-1.25mm]{2.5mm}{2.5mm}}
\drawline\fermion[\NE\REG](\pbackx,\pbacky)[4500]
\drawarrow[\LDIR\ATTIP](\pmidx,\pmidy)
\put(12200,3500){$D$}
\drawline\fermion[\SE\REG](\pfrontx,\pfronty)[4500]
\drawarrow[\LDIR\ATTIP](\pmidx,\pmidy)
\global\advance\pmidy by -1500
\put(12200,-4500){$D$}
%
\drawline\fermion[\E\REG](19000,0)[4500]
\drawarrow[\LDIR\ATTIP](\pmidx,\pmidy)
\global\advance\pmidy by -1500
\put(\pmidx,\pmidy){$B$}
\put(\pbackx,\pbacky){\rule[-1.25mm]{2.5mm}{2.5mm}}
\drawline\fermion[\NE\REG](\pbackx,\pbacky)[4500]
\drawarrow[\LDIR\ATTIP](\pmidx,\pmidy)
\global\advance\pmidx by 2000
\global\advance\pmidy by 1600
\put(\pmidx,\pmidy){$D$}
\drawline\fermion[\E\REG](\pfrontx,\pfronty)[4500]
\drawarrow[\LDIR\ATTIP](\pmidx,\pmidy)
\global\advance\pmidx by -1000
\global\advance\pmidy by -1500
\put(\pmidx,\pmidy){$P_c$}
\put(28000,0){\circle*{800}}
\drawline\scalar[\NE\REG](\pbackx,\pbacky)[2]
\drawarrow[\LDIR\ATTIP](\pmidx,\pmidy)
\put(31200,3500){$\pi$}
\drawline\fermion[\SE\REG](\pfrontx,\pfronty)[4500]
\drawarrow[\LDIR\ATTIP](\pmidx,\pmidy)
\global\advance\pmidy by -1500
\put(31200,-4500){$D$}
\end{picture}
\vskip 2 cm
\hskip -2 cm {\bf (a)} \hskip 7 cm {\bf (b)}
\vskip 1 cm
{ Figure 1: Polar diagrams contributing to ${\overline {\cal A}}_B$ and
${\overline {\cal A}}_R$. The dot corresponds to a strong vertex, the box to a 
weak vertex}
\end{center}
\end{figure}
%
The first set of contributions ${\overline {\cal A}}_B$ includes those
intermediate states which become degenerate in mass with the initial $B$
or the final $D$ states in the HQET: the $J^P=1^-$ $B^*$ (fig. 1a) and
$D^*$ (fig.
1b) mesons. The second set ${\overline {\cal A}}_R$ 
denotes contributions from excited beauty
and charm mesons, corresponding to P-waves in the constituent quark model:
$B_0$, $B^*_2$ and $D_0$, $D^*_2$, with $J^P=(0^+, 2^+)$, respectively.
Clearly, this is  a simplification, since in principle 
the contribution of other
intermediate resonances can be considered, such as, e.g., 
$B \to \psi^{(n)} \pi^0 \to D^+ D^- \pi^0$. In the case of $\psi^{''}$,
which should be the most important one, an estimate of this
colour-suppressed process in the factorization approximation gives a
negligible contribution with respect to the other ones considered here. 

Since $H_W$ 
has a $(V-A)\times (V-A)$ structure, for the equal-time commutator in Eq.
(\ref{ward}) the equality 
$[F_5^3, H_W]=-[F^3, H_W]$ holds, with $F^3$ the isotopic spin operator.
Then, using $F^3|{\bar B}^0>={1 \over 2} |{\bar B}^0>$ and 
$F^3|D^+ D^->_{(I=1,0 ; I_3=0)}=0$, the equal-time commutator  becomes:
\begin{equation}
{\overline {\cal A}}_1=-i {\sqrt{2} \over f_\pi} 
<D^+(p_+) D^-(p_-) |H_W|{\bar B}^0(p)> \;. \label{eq-time}
\end{equation}
The separation indicated in the second and the third terms on 
the right hand side of Eq. (\ref{ward}) is done in order to avoid the
ambiguity which arises in taking the mass degeneracy limit first and then
the limit $q \to 0$ or viceversa in $i q^\lambda M_\lambda$ or 
${\overline {\cal A}}_B$, so that 
$\Big[-i {\sqrt{2} \over f_\pi} q^\lambda M_\lambda-
{\overline {\cal A}}_B \Big]$ has a
well defined limit when $q \to 0$. ${\overline {\cal A}}_B$ and $M_\lambda$ 
(Born) (which alone is relevant for the above limit) can be easily calculated, and as a
result Eq. (\ref{ward}) becomes:
\begin{equation}
{\overline {\cal A}}(s_+,s_-)=
{\overline {\cal A}}_1(s_+,s_-)+{\overline {\cal A}}_2^{B^*}(s_+,s_-)+
{\overline {\cal A}}_4^{D^{*-}}(s_+,s_-)
+{\overline {\cal A}}_5^{D^{*+}}(s_+,s_-)+{\overline {\cal A}}_R \;, 
\label{sum}
\end{equation}
\noindent where
\begin{eqnarray}
{\overline {\cal A}}_2(s_+,s_-)&=
& g_{B^{*0}B^0\pi^0}\Big[{1 \over m_B^2} \Big(1+{m_{B^*}^2
\over m_{B^*}^2-s_0} \Big) (p-q)_\mu-{ 1\over m_{B^*}^2-s_0}(p+q)_\mu
\Big] F^\mu \label{a2r} \\
{\overline {\cal A}}_{4,5}(s_+,s_-)&=
& g_{D^{*\mp}D^\mp\pi^0}\Big[-{1 \over m_D^2}
\Big(1+{m_{D^*}^2 
\over m_{D^*}^2-s_\mp} \Big) (p_\pm+q)_\mu+{ 1\over
m_{D^*}^2-s_\mp}(p_\pm-q)_\mu
\Big] G^\mu_\pm \label {a45r}
\end{eqnarray}
with
\begin{eqnarray}
F^\mu \epsilon_\mu &=& <D^+ D^-|H_W|{\bar B}^{*0}> \nonumber \\
G^\mu_\pm \eta_\mu^\pm &=& <D^\pm D^*|H_W|{\bar B}^0> \;\;\;,\label{fg}
\end{eqnarray}
$\epsilon$ and $\eta$ being the $B^*$ and $D^*$ polarization vectors, 
respectively.
It may be noted that the constant terms in Eqs. (\ref{a2r}) and
(\ref{a45r}) correspond to the limit $q \to 0$ indicated in Eq.
(\ref{ward}). Here, consistently with the use of the infinite heavy quark mass 
limit,  we have neglected terms of order
$\delta_B={m_{B^*}^2-m_B^2 \over 2 m_B}$ and $\delta_D$
in comparison with the heavy meson masses. 

With $H_W$ given
in Eq. (\ref{eq:hamilt}) and using the factorization ansatz, the matrix
elements 
(\ref{eq-time}) and (\ref{fg}) can be evaluated in 
the heavy quark effective theory (HQET) in terms of the
Isgur-Wise semileptonic form factor $\xi(v \cdot v^\prime)$, where $v$ and
$v^\prime$ are the relevant four-velocities.
As for the polar terms in ${\overline {\cal A}}_R$ in Eq. (\ref{sum}), we only 
consider $P$-wave intermediate charm and beauty resonances. In this case,
the matrix elements corresponding to the relevant amplitudes can be written in 
terms of the Isgur-Wise universal form factors
$\tau_{1/2}(v \cdot v^\prime)$ and $\tau_{3/2}(v \cdot v^\prime)$. Defining 
$\displaystyle{{\cal K}={G_F \over \sqrt{2}}(c_1 + {c_2 \over N_c})
V_{cb} V_{cd}^*}$, and
parameterizing the effective strong couplings
\begin{eqnarray}
g_{B^{*+}B^0\pi^+}&=& {1 \over \sqrt{2}} g_{B^{*0}B^0\pi^0}=
{2 m_B \over f_\pi} g \nonumber \\
g_{D^{*-}D^0\pi^-}&=& {1 \over \sqrt{2}} g_{D^{*-}D^-\pi^0}=
{2 m_D \over f_\pi} g \;, \label{gbbpi}
\end{eqnarray}
\noindent and the current-particle vacuum matrix elements
\begin{eqnarray}
<D^-(p_-)|{\bar c}\gamma^\mu (1-\gamma_5)d|0>= i {\hat F \over \sqrt{m_D}} 
p_-^\mu \nonumber \\
<D^-_0(k)|{\bar c}\gamma^\mu (1-\gamma_5)d|0>= - i {\hat F^+ \over
\sqrt{m_{D_0}}} k^\mu \;, \label{fhat}
\end{eqnarray}
\noindent we can list the expressions of the various contributions introduced
above. The equal time contribution is simply given by:
\be
{\overline {\cal A}}_1(s_+,s_-)=
- {{\cal K} \hat F \sqrt{m_B} \over \sqrt{2} f_\pi}  
\xi({m_B^2+m_D^2-s_-\over 2 m_B m_D}) 
\Big[{m_B^2 +m_\pi^2-s_+-s_-\over 2 m_D} +{m_D^2 +m_B^2-s_+\over 2 m_B} \Big]
\;\;\;. \label{eq:etime}
\ee
As for the polar terms, the contribution of the $B^*$ intermediate particle 
in fig.1a is: 
\bea
{\overline {\cal A}}_2(s_+,s_-)&=& {\sqrt{2} {\cal K} \hat F g \over f_\pi}
{\sqrt{m_B} m_{B^*} \over s_0-m_{B^*}^2} \xi({s_0 \over 2 m_D m_{B^*}})
\Big[ -(1+ {s_0\over 2 m_D m_{B^*}}) {s_--m_D^2-m_\pi^2\over 2 } \nn\\
&+& {s_0\over 2 m_{B^*}}{s_+-m_D^2-m_\pi^2\over 2 m_D}
+{s_0 \over 4 m_{B^*}^2}(m_B^2-m_\pi^2-s_0) \Big] 
\label{eq:a2}
\eea
(we neglect the $B^*$ width); on the other hand, 
the contribution of the $B_0$, $J^P=0^+$ state, 
whose width is $\Gamma_b$, reads:
\be
{\overline {\cal A}}_3(s_+,s_-)= 
{{\cal K} \hat F G_{B_0 B \pi}(s_0) \over s_0-m_{B_0}^2+i m_{B_0} \Gamma_b}
{1 \over 2 m_D \sqrt{m_{B_0}}} \tau_{1/2}( {s_0\over 2 m_{B_0} m_D})
\big[ (m_{B_0}-m_D)s_0-2 m_D^2 m_{B_0}\big] \;\;\;.
\ee
The contributions of the poles $D^{*-}$ and $D^{*+}$ in fig.1b
($\Gamma_{D^*}$ is the  $D^{*}$ width) are, respectively: 
\bea
{\overline {\cal A}}_4(s_+,s_-)&=& 
{ \sqrt{2} {\cal K} \hat F g \over f_\pi} 
{\sqrt{m_B} m_D m_{D^*} \over s_--m_{D^*}^2+i m_{D^*}\Gamma_{D^*} }
\xi({m_B^2 + m_D^2 -s_- \over 2 m_B m_D}) 
\Big( {m_B + m_D \over m_B m_D} \Big)
\Big[ - {s_+ + s_- -2 m_D^2 \over 2} \nn\\
&+& {s_--m_D^2+m_\pi^2 \over 2 m_{D^*}^2} {m_B^2-m_D^2+s_- \over 2} \Big]
\;\;\; ,
\eea
\bea
{\overline {\cal A}}_5(s_+,s_-)&=& 
-{ \sqrt{2} {\cal K} \hat F g \over f_\pi} {1 \over 4 \sqrt{m_B}}
{1 \over s_+-m_{D^*}^2+i m_{D^*}\Gamma_{D^*} }
\xi({m_B^2 - m_D^2 +s_+ \over 2 m_B m_{D^*}}) \nn\\
&\Big[& -(s_+ + s_- -2 m_D^2) (m_B^2 - m_D^2 - s_+) 
+(2 m_B m_{D^*} + m_B^2 - m_D^2 + s_+)(s_--m_D^2-m_\pi^2) \nn\\
&-&{m_B \over m_{D^*}} (m_B^2-m_D^2-s_+)(m_\pi^2-m_D^2+s_+) \Big] 
\;\;\; .
\eea
 The contributions of $D^{-}_0$ and $D^{+}_0$ in fig.1b can be written 
as
\be
{\overline {\cal A}}_6(s_+,s_-)=
- { {\cal K} G_{D_0 D \pi}(s_-) \hat F^+ \over 2 \sqrt{m_{D_0} m_B m_D}}
{1 \over s_- - m_{D_0}^2 + i m_{D_0} \Gamma_{D_0}}
\xi({m_B^2 + m_D^2 -s_- \over 2 m_B m_D}) (m_B-m_D) [(m_B+m_D)^2-s_-]
\ee
and
\bea
{\overline {\cal A}}_7(s_+,s_-)&=&
{ {\cal K} G_{D_0 D \pi}(s_+) \hat F \over 2 \sqrt{m_{D_0} m_B m_D}}
{1 \over s_+ - m_{D_0}^2 + i m_{D_0} \Gamma_{D_0}}
\tau_{1/2}({m_B^2 - m_D^2 +s_+ \over 2 m_B m_{D_0}}) 
\Big[ m_{D_0} (m_B^2+m_D^2-s_+) \nn\\
&-&m_B (m_B^2-m_D^2-s_+) \Big]
\eea
respectively. In the previous equations, the following definition holds:
\be
G_{D_0 D \pi}(s)=-\sqrt{m_{D_0} m_D\over 2}
{s-m_D^2\over m_{D_0}} {h\over f_\pi} \;\;\;,
\ee
and an analogous expression is used for $G_{B_0 B \pi}$.

Finally, we consider the contribution of the $ D^{*+}_2$  pole, 
whose width is $\Gamma_{D_2}$:
\bea
{\overline {\cal A}}_8(\ssp,\sm) &=&
{\cal K} {\hat F} \sqrt{6}  {h^\prime \over f_\pi}  
\tau_{3/2}( {s_+ + m_B^2 - m_D^2 \over 2 m_B m_{D_2}})
{\sqrt{m_B} m_{D_2} (m_B+m_{D_2}) \over s_+ - m_{D_2}^2 + i m_{D_2} 
\Gamma_{D_2}} \nonumber \\
&&\Big\{ \Big[ {s_+ + s_- - 2 m_D^2 \over 2 m_B} 
- {s_+ + m_B^2 - m_D^2 \over 2 m_B m_{D_2}} 
{s_+ - m_D^2 + m_\pi^2 \over 2 m_{D_2}}  \Big]^2 \nonumber \\
&-& {1 \over 3} \Big[1-({s_+ + m_B^2 -  m_D^2 \over 2 m_B m_{D_2}})^2  \Big] 
\Big[ m_\pi^2 - ({s_+ -  m_D^2 + m_\pi^2 \over 2 m_{D_2}})^2  \Big]\Big\} 
\;\;\;,
\eea
and the contribution of the $B^{*0}_2$ intermediate state:
\bea
{\overline {\cal A}}_9(\ssp,\sm) &=&
-{\cal K} {\hat F} \sqrt{6}  {h^\prime \over f_\pi}  
\tau_{3/2}( {s_0 \over 2 m_B m_{B_2}})
{\sqrt{m_B} m_{B_2} (m_B+m_{D_2}) \over s_0 - m_{B_2}^2 + i m_{B_2} 
\Gamma_{B_2}} \nonumber \\
&&\Big\{ \Big[ {s_+ -  m_D^2 - m_\pi^2 \over 2 m_D} 
- {s_0 \over 2 m_B m_D}  {s_+ + s_- - 2 m_D^2 -2 m_\pi^2  \over 2 m_{B_2}}
\Big ]^2 \nonumber \\
&-&  {1 \over 3} \Big[1-{s_0^2 \over 4 m_B^2 m_{B_2}^2}  \Big] 
\Big[ m_\pi^2 - ({s_+ +s_- - 2 m_D^2 -2 m_\pi^2 \over 2 m_{B_2}})^2 \Big]\Big\} 
\;\;\;. 
\label{eq:a9}
\eea
The $D_2^{*-}$ contribution vanishes in the factorization approximation.
Notice that, for simplicity, we have assumed
momentum-independent widths in the Breit-Wigner denominators. 

In the above equations, the usual definitions of the universal Isgur-Wise form 
factors have been used (see, e.g., the reviews 
\cite{review,report}); 
as for the the effective coupling $h^\prime$ in the 
$D^*_2 D \pi$ and $B^*_2 B \pi$ vertices, it has been first
investigated in \cite{falk} in the framework of HQET and
we shall turn to this coupling in the next Section. 

Eqs.(\ref{eq:a2})-(\ref{eq:a9}) are obtained by considering the expressions
for the effective strong vertices and the weak ones in the factorization
approximation,
 and combining them to evaluate the diagrams in fig.1a,b. This 
procedure presents some uncertainties, for example related to the relative
signs 
between the various contributions. A method which allows to partially 
overcome 
such difficulties is based on the use of an effective heavy meson chiral 
Lagrangian, and the next Section is devoted to this approach.

\section{Evaluation by an effective chiral Lagrangian}\label{s:efflag}
In order to determine an expression for the amplitude (\ref{process})
let us consider the effective Lagrangian \cite{report}
\begin{equation}
{\cal L}=i g {\mathrm {Tr}}(\overline{H} H \gamma^{\mu}\gamma_5
{\cal A}_\mu ) + \left(i h 
{\mathrm {Tr}}(\overline{H} S \gamma^{\mu}\gamma_5
{\cal A}_\mu ) +\frac{i}{\Lambda_\chi} 
{\mathrm {Tr}}(\overline{H} T^\mu \gamma^{\lambda}\gamma_5 [
h_1 D_\mu {\cal A}_\lambda + h_2 D_\lambda {\cal A}_\mu]) +h.c.\right),
\label{strong}
\end{equation}
that describes  the strong interactions of pions and kaons with
 heavy mesons containing one heavy quark.
This construction of the effective vertices follows the 
prescription of HQET,  
with the further constraints imposed by chiral symmetry. 
$H$, $S$ and $T$ represent heavy meson doublets corresponding to different
values of the spin-parity $s_\ell^P$ of the light degrees of freedom 
of a $\bar q Q$ meson; 
the doublet $H$ comprises  the negative parity 
low lying states, viz. $D,~D^*$ in the case of  
charm and $B,~B^*$ for beauty; the multiplet $S$ 
is characterized by $s_\ell^P={1 \over 2}^+$ and comprises  the positive
parity ($J^P=0^+,~1^+$) low lying states, viz. $D_0,~D_1^{*\prime}$ for charm 
and $B_0,~B_1^{*\prime}$ for beauty;
the multiplet $T$ has $s_\ell^P={3 \over 2}^+$ and comprises  
the positive parity ($J^P=1^+,~2^+$) 
states:  $D_1^{*},~D^*_2$ for charm and $B_1^{*},~B^*_2$ for beauty.

The fields $H$ and $S$ and $T$ are $4\times4$ matrices 
containing annihilation operators.  In the  charm sector,
for the negative parity states $s_\ell^P={1 \over 2}^-$
 these fields are given by
\begin{equation}
H = \frac{(1+\slash v)}{2}\; [D_{\mu}^*\gamma^\mu - D \gamma_5 ]\;\;\;,
\label{eq:fieldh}
\end{equation}
and the conjugate field is given by ${\bar H} = \gamma_0 H^{\dagger}
\gamma_0$. For positive 
parity $s_\ell^P=({1 \over 2}^+,{3 \over 2}^+)$ states, the fields are
defined by
\begin{eqnarray}
S &=&{{1+\slash v}\over 2}[D_{1\mu}^{*\prime} \gamma^\mu\gamma_5-D_0]\\
T^\mu & =& {\frac {(1+ \slash v)}{2}}
\left[D_2^{*\mu\nu}\gamma_\nu-\sqrt{\frac 3 
2} D^*_{1\nu}\gamma_5
\left(g^{\mu \nu}-\frac 1 3 \gamma^\nu(\gamma^\mu-v^\mu)\right)\right].
\label{eq:fieldt}
\end{eqnarray}
In Eqs.(\ref{eq:fieldh})-(\ref{eq:fieldt})
$v$ generically represents the heavy meson four-velocity, 
 $D^{*\mu}$, $D$, $D_{1\mu}^{*\prime}$ and $D_0$ are annihilation
operators normalized as follows:
\begin{eqnarray}
\langle 0|D| c{\bar q} (0^-)\rangle & =&\sqrt{M_H} \label{eq:34}\\
\langle 0|{D^*}^\mu| c{\bar q} (1^-)\rangle & = & \epsilon^{\mu}\sqrt{M_H}
\;\;\;, \label{eq:35}
\end{eqnarray}
and similar equations hold for the positive parity states
(in Eqs. (\ref{eq:34}) and (\ref{eq:35}) $M_H=M_D=M_{D^*}$
is the common mass in the $H$ multiplet);
the transversality conditions are
$v^\mu D^*_{\mu}=v^\mu D_{1\mu}^{*\prime}=v^\mu D^*_{1\mu}=
v^\mu D^*_{2\mu\nu}= 0$.

The couplings $HH\pi$,  $HS\pi$ and $H T \pi$ of the heavy mesons
with light pseudoscalar mesons are constructed 
through  the axial vector current
\be
{\cal A}^\mu = {i\over 2} (\xi^\dagger \partial^\mu \xi -\xi \partial^\mu
\xi^\dagger)~, \label{av}
\ee
where $\xi=\exp(i \Pi/f_\pi )$, with
 $\Pi$ the familiar $3\times 3~SU(3)$
matrix describing the octet of light pseudoscalar mesons. 
As it is clear from Eq.(\ref{strong}), the interaction vertices 
$HH\pi$,  $HS\pi$ and $H T \pi$ are described in terms of the effective 
couplings $g$, $h$ and $h_{1,2}$. We shall quote in the next Section the 
numerical values for such parameters; here, we only notice that in our 
calculation the combination
\be
h^\prime=\frac{h_1+h_2}{\Lambda_\chi}
\ee
($\Lambda_\chi$ is a mass parameter) is needed, which can be determined from 
the experimental measurement of the $D_2^*$ pionic transitions.

In terms of the heavy and light meson operators,  
an effective weak nonleptonic Lagrangian can be written as follows:
\be
{\cal L}_{eff}=\frac{G_F}{\sqrt{2}} ~(c_1+ {c_2\over N_c})~ V_{cb}~ V_{cd}^*~
{\mathrm {Tr}}[(L^\mu+ L^{\prime \mu})_{({\bar c})}
(J_\mu+ J_\mu^{\prime})_{(cb)}]\;\;\;.
\label{eq:effweak}
\ee
The effective currents $L^\mu_{({\bar c})}
,~ L^{\prime \mu}_{({\bar c})}
$ (the subscript ${\bar c}$ means anti-charm) are
\begin{eqnarray}
L^\mu_{({\bar c})} &=&\frac{i}{2} ~{\hat F}~{\mathrm {Tr}}(\gamma^\mu (1-\gamma_5) 
H_{({\bar c})} \xi^\dagger)\\
L^{\prime\mu}_{({\bar c})} &=&\frac{i}{2}~ {\hat F}^+ {\mathrm {Tr}}(\gamma^\mu (1-
\gamma_5) S_{({\bar c})} \xi^\dagger)
\end{eqnarray}
with $\hat F$ and ${\hat F}^{+}$ 
already introduced in (\ref{fhat}).
The effective $ J_\mu$ and $J_\mu^{\prime}$ currents 
describing the weak $b\to c$ transition can be written in terms of universal
form factors:
\begin{eqnarray}
J^\mu_{(cb)} &=
&-\xi(v \cdot v^\prime)~{\mathrm {Tr}}({\bar H}_{(c)}(v^\prime)
\gamma^\mu (1-\gamma_5) H_{(b)}(v)) \\
J^{\prime\mu}_{(cb)} &=
&-\tau_{1/2}~(v\cdot v^\prime){\mathrm {Tr}}({\bar S}_{(c)} 
(v^\prime)\gamma^\mu (1-\gamma_5) H_{(b)}(v)) \\
 J^{\prime \prime \mu}_{(cb)} &=&- \sqrt{3} \tau_{3/2}~(v \cdot v^\prime )
{\mathrm {Tr}}(v_\lambda
{\bar T}^\lambda_{(c)}(v^\prime) \gamma^\mu (1-\gamma_5) H_{(b)}(v))
\end{eqnarray}
where $v^\mu,~v^{\prime\mu}$ are the heavy meson four-velocities
in the initial and final state. 

The Lagrangian (\ref{eq:effweak}) meets the
following 
requirements: 
i) it allows 
$b \to c \bar c d$ transitions with $B$ (or $B^*$) in the initial state, 
and two charmed mesons (with $s_\ell^P = {1\over 2}^-, {1\over 2}^+$ or
${3\over 2}^+$) plus any number of pseudoscalar light mesons in the final
state;
ii) the resulting amplitude corresponds to the evaluation of the weak
4-quark effective nonleptonic Lagrangian in the factorization approximation;
iii) it contains the minimum   number of light meson field derivatives, 
consistently with general properties of chiral symmetry and the soft-pion
limit procedure 
employed in the previous Section. 

The effective weak nonleptonic Lagrangian (\ref{eq:effweak}), together with 
the strong interaction Lagrangian (\ref{strong}), allows to write down an 
expression for the set of amplitudes contributing to the transition 
(\ref{process}).
The equal-time contribution derived from 
Eqs.(\ref{strong},\ref{eq:effweak}):
\be
{\overline {\cal A}}_1(\ssp,\sm)=
- \frac{{\cal K}{\sqrt{m_B}} m_D}{{\sqrt 2}f_\pi}   ~ \fhat ~ 
\xi(v \cdot v_+)( v_+ \cdot v_- + v \cdot v_-) 
\ee                                                                      
exactly reproduces Eq.(\ref{eq:etime})
taking into account the invariants in Eq. (\ref{eq:invariants}).
Together with this term, the set of polar contributions corresponding to
Eqs.(\ref{eq:a2})-(\ref{eq:a9}) can be written;
the differences with respect to the expressions reported in the previous 
Section represent a set of finite mass corrections,
that  partially account for the theoretical uncertainties
in the calculation. They correspond to different treatments of  
 the Breit-Wigner forms.
 
The $B^{* }$ and  $B_0$ (of width $\Gamma_b$) pole  contributions 
are given by
\bea
{\overline {\cal A}}_2(\ssp,\sm)&=
&- \frac{{\cal K}{\sqrt{m_B}} m_D}{{\sqrt 2}f_\pi} ~ \fhat~ g~ 
\frac{ \xi(v \cdot v_+)}{\qvp+(\dmb)} \nonumber \\
&(&\vpw~ ( \qv-\vvp   ~ \qvp)-( 1+\vvp ) ( \qw~ -~ \vpw ~\qvp)) 
\eea
and
\be
{\overline {\cal A}}_3(\ssp,\sm)= 
\frac{{\cal K}{\sqrt{m_B}} m_D}{{\sqrt 2}f_\pi} ~ h
~\fhat~\tau_{1/2}(\vvp)~\frac{ \qv~(-\vw+\vpw)}{-\qvp-\dm+i\frac{\Gamma_b}{2}}
\;\;\;,
\ee
respectively,
with $\dm$ the mass difference $\dm=M_{B_0}-M_B=M_{D_0}-M_D$. As for the
$D^{*-}$ and $D^{*+}$ contributions ($\eps$ is the  $D^*$ width),
they read respectively: 
\bea
{\overline {\cal A}}_4(\ssp,\sm)&=&
- \frac{{\cal K}{\sqrt{m_B}} m_D}{{\sqrt 2}f_\pi}  ~g ~ \fhat 
~\xi(\vvp)~\frac{\qv+\qvp-\qw(\vw+\vpw)}{\qw-(\dmd)+i\frac{\eps}{2}} \;\;\;, \\
{\overline {\cal A}}_5(\ssp,\sm)&=
&- \frac{{\cal K}{\sqrt{m_B}} m_D}{{\sqrt 2}f_\pi} 
~\fhat ~g~ \frac{\xi(\vvp)}{\qv-(\dmd)+i\frac{\eps}{2}} \nonumber \\
&[&(1+\vvp)(\qw~-~\vw~ \qv)- \vw~(\qvp~-~\vvp ~\qv)] \;\;\;.
\eea
The $D^{-}_0$ and $ D^{+}_0$ $J^P=0^+$ pole terms are
\be
{\overline {\cal A}}_6(\ssp,\sm)=
\frac{{\cal K}{\sqrt{m_B}} m_D}{{\sqrt 2}f_\pi}  ~h~ \fhatpiu 
~\xi(\vvp) ~\frac{\qw~(\vw+\vpw )}{\qw-\dm+i\frac{ \Gamma_d}{2}}
\ee
and
\be
{\overline {\cal A}}_7(\ssp,\sm)=
\frac{{\cal K}{\sqrt{m_B}} m_D}{{\sqrt 2}f_\pi} ~ h ~\fhat 
~\tau_{1/2}(\vvp)~\frac{\qv~(\vw-\vpw)}{\qv-\dm+i\frac{\Gamma_d}{2}} \;\;\;.
\ee
The $D^{*+}_2$ pole contribution reads:
\bea
{\overline {\cal A}}_8(\ssp,\sm) &=&
\frac{{\cal K}{\sqrt{m_B}} m_D}{{\sqrt 2}f_\pi} ~ \frac{\sqrt{3} h^\prime
\fhat}{\qv-\dm_2+i\frac{\Gamma_2}{2}} ~\tau_{3/2}(\vvp) \nonumber\\
&\times&\Big( - \frac{\vvp +1}{3}[m_\pi^2-(\qv)^2](-\vw+\vpw) \nonumber \\
&+&[\qvp-(\vvp)(\qv)][(\vvp +1)\qw-\vw(\qv+\qvp)]\Big) \;\;\;,
\eea
with $\dm_2=M_{D^*_2}-M_D=M_{B^*_2}-M_B$.
 Finally, the $B^{*0}_2$ pole  contribution is
\bea
{\overline {\cal A}}_9(\ssp,\sm) &=&
\frac{{\cal K}{\sqrt{m_B}} m_D}{{\sqrt 2}f_\pi} ~ \frac{\sqrt{3} h^\prime
 \fhat}{-\qvp-\dm_2+i\frac{\Gamma_2}{2}}~\tau_{3/2}(\vvp) \nonumber\\
&\times&\Big(-\frac{\vvp +1}{3}[m_\pi^2-(\qvp)^2](\vw-\vpw) \nonumber \\
&-&[\qv-(\vvp)(\qvp)][(\vvp +1)\qw-\vpw(\qv+\qvp)]\Big).
\eea
The expressions for the various terms 
${\overline {\cal A}}_1-{\overline {\cal A}}_9$ allow us to reconstruct the 
amplitude describing  the process (\ref{process}). The
differences with respect to the results obtained from the amplitude derived 
in the previous Section  represent a theoretical uncertainty
associated to the writing of the polar contributions.

\section{Numerical analysis}\label{s:res}

In order to estimate  the rate
of the decay (\ref{process}) and the terms in Eq.(\ref{eq:gis}),
using the formulae in the previous Sections,
we must rely on numerical values for the various hadronic parameters such as 
the leptonic constants, the strong couplings and the semileptonic form 
factors. In some cases, experimental information can be used; 
theoretical methods can be adopted to determine the remaining quantities,
and for these we shall mainly use the results of the QCD sum rule 
method. 

The strong coupling constants appearing in (\ref{strong}) 
have been evaluated by several authors \cite{sr1,lcsr};
we  assume here  for $g$ and $h$ the values given in 
Ref. \cite{sr1}: $g=0.35,~h=-0.52$. 
The combination 
$h^\prime=\frac{h_1+h_2}{\Lambda_\chi}$
can be obtained from experimental data; from  the full width of the 
$D^*_2(2460)$ state, $\Gamma_2=23\pm 5$ MeV \cite{PDG},
we get $h^\prime=0.60$ GeV$^{-1}$, assuming that $D^*_2 \to D \pi$
saturates the hadronic $D^*_2$ width.

The constants ${\hat F},~{\hat F}^+$ do not depend on the heavy
quark mass (modulo logarithmic corrections) and have been 
estimated by QCD sum rules \cite{sr1,broad,dai}. 
We take the values: 
$\fhat=0.30$ GeV$^{3/2}$ and $\fhatpiu =0.46$ GeV$^{3/2}$, 
corresponding to the results at zero order in the strong coupling $\alpha_s$.
It should be noticed that for some of these parameters, as well as for the 
Isgur-Wise form factors $\xi$ and $\tau_{1/2}$, 
the ${\cal O}(\alpha_s)$ corrections have been computed 
\cite{broad,neubertxi,bagan,newtau}. 
However, since such corrections are not known for all the parameters needed 
in the present calculation, for consistency we  use  the values 
obtained at zero order in $\alpha_s$, including the effects of the 
known radiative corrections in the estimate of the theoretical uncertainty of 
our results. 

The universal form factors $\xi$, $\tau_{1/2}$ and
$\tau_{3/2}$ can be parametrized 
as follows:
\bea
\xi(\omega)&=&(\frac{2}{1+\om})^2 \label{xi}\\
\tau_{1/2}(\om)&=&0.3~[1-0.5 (\om-1)]\\
\tau_{3/2}(\om)&=&0.3~[1-0.8 (\om-1)] \;\;\;.
\eea
Eq.(\ref{xi}) is a useful parameterization of the Isgur-Wise form factor,
obeying the normalization condition $\xi(1)=1$ dictated by the heavy quark
symmetry, and having a slope compatible with experimental data.
The parameterizations for the $\tau_i$ can be found,
e.g., in \cite{review,oldtau}, taking into account an
uncertainty of $15 \%$ for the value at the zero recoil point $\om=1$.
\cite{dai1}

Besides the already mentioned values for the strong coupling constants 
we use the following numerical
values for the physical parameters appearing in the previous formulae: 
$\dmb=0.045 $ GeV and $\dmd=0.142 $ GeV, $V_{cb}=0.04$ and 
$V_{cd}=0.22$ \cite{PDG}; $c_1\simeq 1.2$ and $c_2=-0.2$ \cite{review}.
As for the mass difference $\dm_2=m_{D^*_2}-m_D$, we use $\dm_2=500 $ MeV
\cite{PDG}
and the same value for $\dm_2=m_{B^*_2}-m_B$.
For the other quantities we use the theoretical determinations 
$\dm=m_{B_0}-m_B\simeq 0.50$ GeV, $\Gamma_b=0.30 $ GeV,
$\Gamma_d=0.14 $ GeV,  
$\eps=35 $ KeV. 
These values
for the $D^*$, $D_0$ and $B_0$ widths are consistent with the values for
the strong coupling constants $g$ and $h$ given above 
(for a discussion see Ref.\cite{sr1}). 

The decay width is given by 
\be
\Gamma({\bar B }^0 \to D^+ D^- \pi^0)=\int_{(m_D+m_\pi)^2}^{(m_B-m_D)^2}ds_-
\int_{s_+^{min}}^{s_+^{max}} ds_+ \frac{d\Gamma}{ds_+ ds_-}
\ee
where
\begin{equation}
\frac{d\Gamma}{ds_+ ds_-}({\bar B }^0 \to D^+ D^- \pi^0)=
{1 \over (2 \pi)^3 } {1 \over 32 m_B^3} |{\overline {\cal A}}|^2 
\end{equation}
and
\begin{equation}
{\overline {\cal A}} =\sum_{k=1}^{9}{\overline {\cal A}}_k \label{a} \;\;\;.
\end{equation}
Using the above parameters, and the formulae reported in the previous Sections,
we find
\be
\Gamma({\bar B }^0 \to D^+ D^- \pi^0) \simeq 5\times 10^{-16}{\rm GeV}
\ee
and the corresponding  branching ratio 
${\cal B}({\bar B }^0 \to D^+ D^- \pi^0)\simeq 1\times 10^{-3}$. Although it is 
difficult to assess the theoretical uncertainty related to this result,
our calculation suggests that a relevant signal of the Cabibbo-suppressed
$B \to D^+ D^- \pi^0$ decay should be detected at the B-factories.
\footnote{Another Cabibbo suppressed $B$ decay to charm mesons has been 
recently observed by the CLEO II Collaboration; it is the process
$B \to D^{*+} D^{*-}$, with a measured branching ratio
${\cal B}(B  \to D^{*+} D^{*-})=[6.2 ^{+ 4.0}_{- 2.9} \pm 1.0] \times 10^{-4}$.
\cite{cleo98}}

As for the size of the various terms appearing in the amplitude 
$\overline{\cal A}$, 
the $B^*$-type resonances in fig.1a give a negligible contribution to the 
final result, whereas the contribution of the charmed intermediate states is 
dominant. Regarding  the equal-time contribution, by itself this term
 would give a width 
of about $7 \times 10^{-17}$ GeV for process (\ref{process}).
Thus, although not quite dominant, it represents 
a contribution to the decay rate whose size is comparable to that of the 
main resonant terms.
Moreover, its presence implies the  
specific dependence of the amplitude on $s_+$ and $s_-$ that  
takes  the chiral symmetry constraint into account.
 
\begin{figure}[h]
\begin{center}
\begin{tabular}{cc}
\epsfig{file=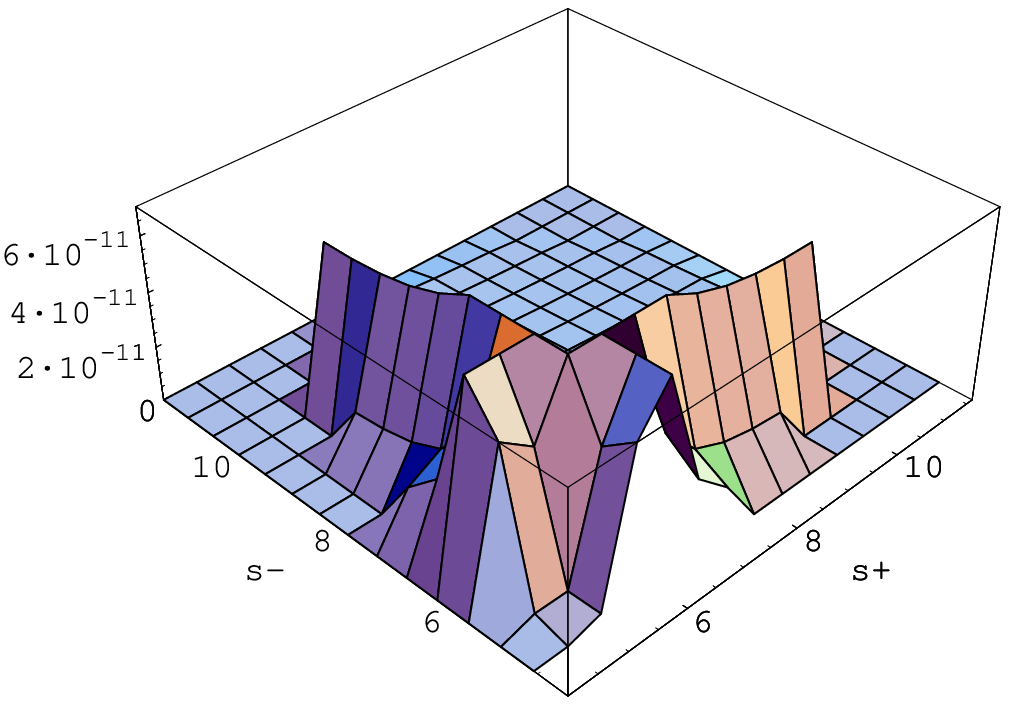,height=6cm} &
\epsfig{file=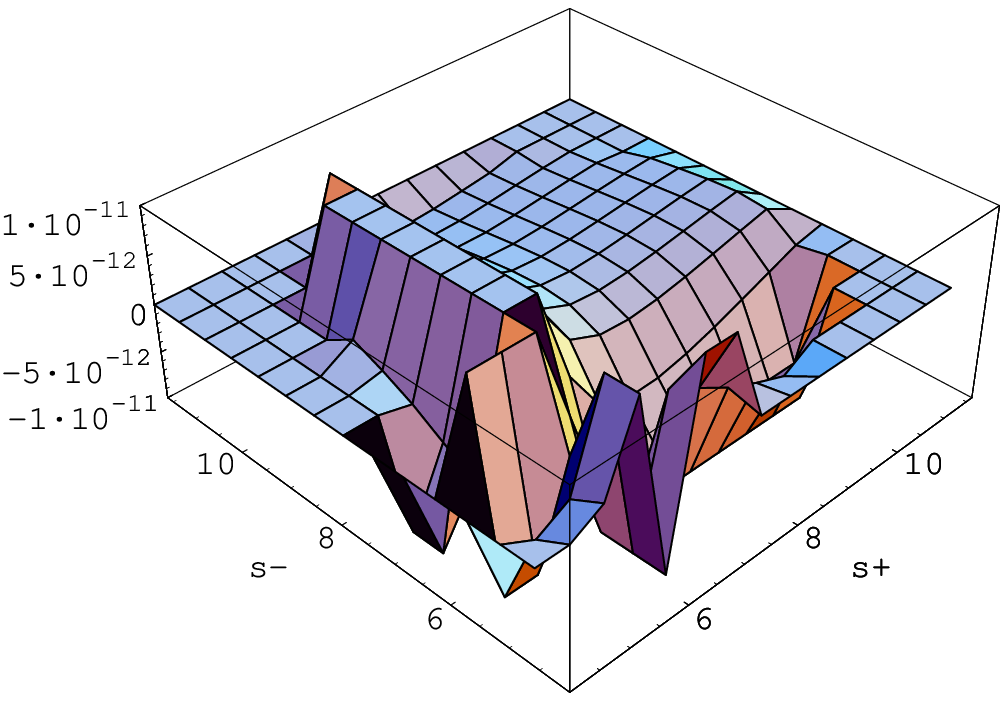,height=6cm} \\
\epsfig{file=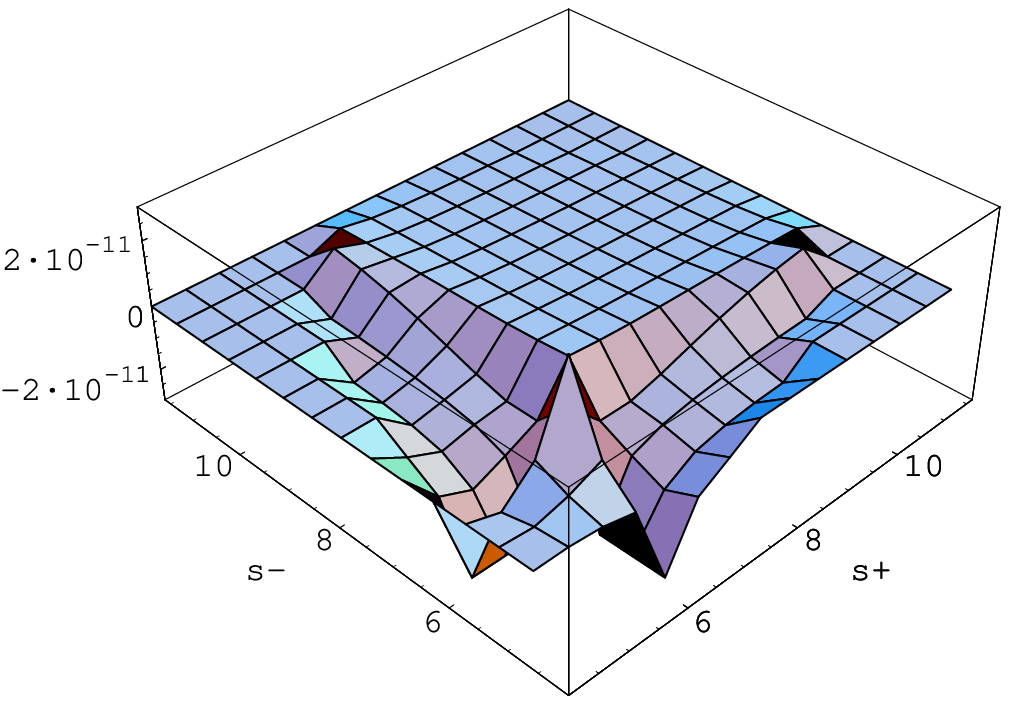,height=6cm} &
\epsfig{file=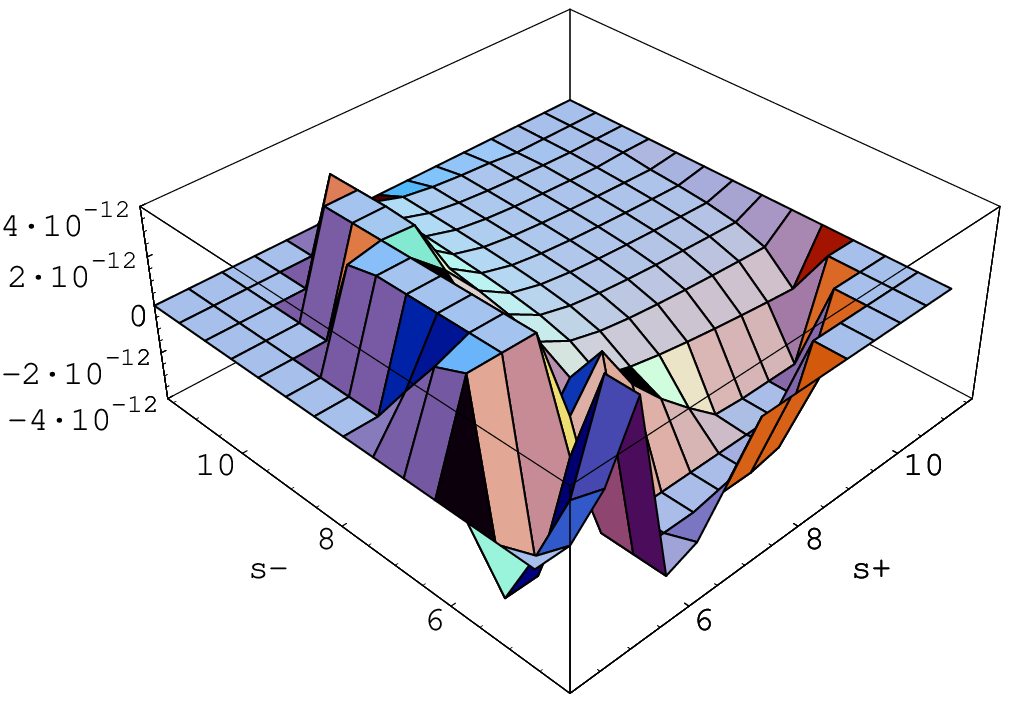,height=6cm} 
\end{tabular}
\label{f:fig2}
\caption{The functions $G_0$ (up-left), $G_c$ (up-right), 
$Re \tilde G_s$ (down-left) and $Im \tilde G_s$ (down-right) in 
Eq.(\ref{eq:gis}) using the amplitudes reported in Section \ref{s:low}.
The variables $s_\pm$ are in $GeV^2$.}
\end{center}
\end{figure}

We now consider the time-dependent decay probabilities (\ref{eq:at}) and 
(\ref{eq:at1}), which are given in terms of the functions
$G_0$, $G_c$ and $\tilde G_s$ in Eq.(\ref{eq:gis}). Using the  
amplitudes ${\overline {\cal A}}$ derived in Section \ref{s:low} 
and the values of the input parameters 
listed
above, we get the functions depicted in  fig. 2. 
On the other hand, the functions obtained by
the parameterization of the amplitude using the 
effective Lagrangian in Section \ref{s:efflag} are depicted in fig. 3;
the comparison between the two figures shows some
agreement between the two methods of extrapolating the polar terms. 
\begin{figure}[h]
\begin{center}
\begin{tabular}{cc}
\epsfig{file=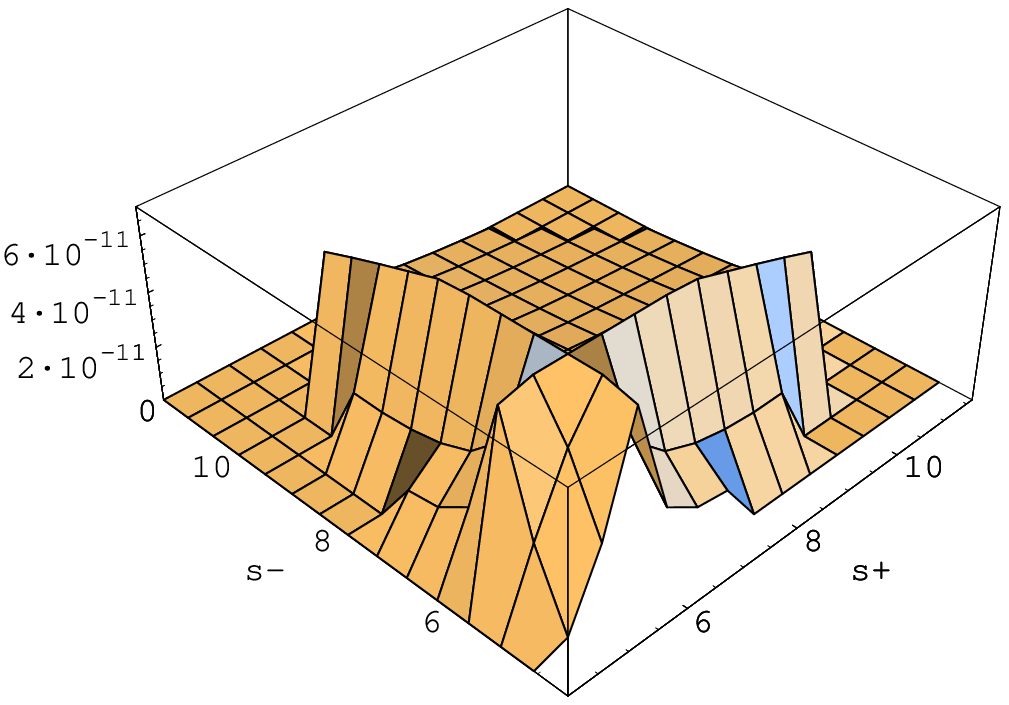,height=6cm} &
\epsfig{file=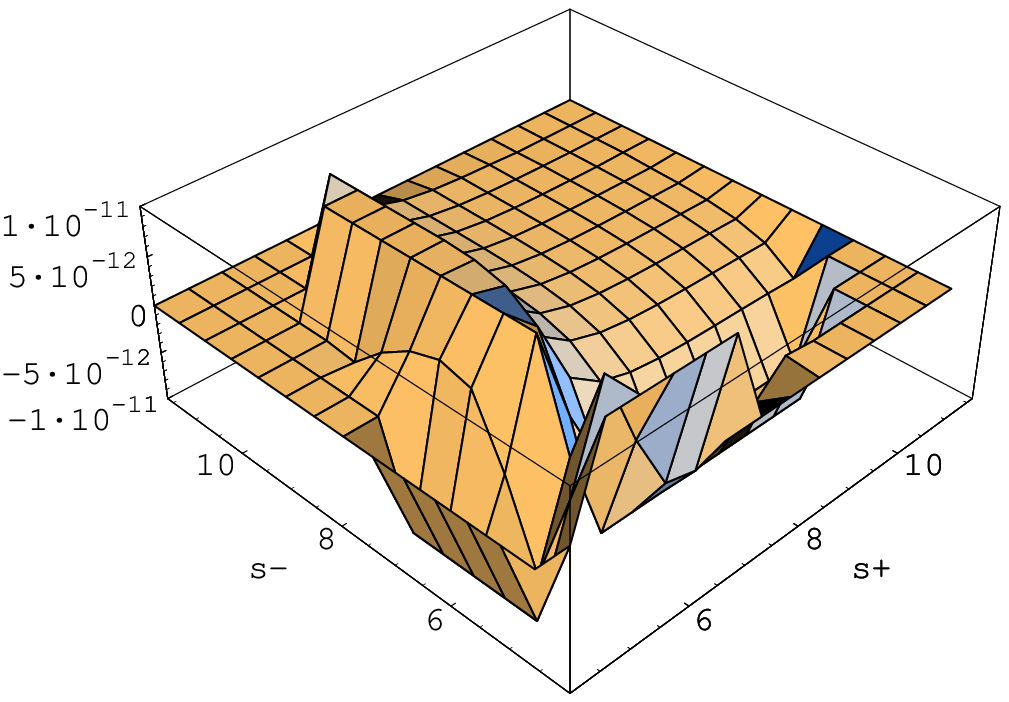,height=6cm} \\
\epsfig{file=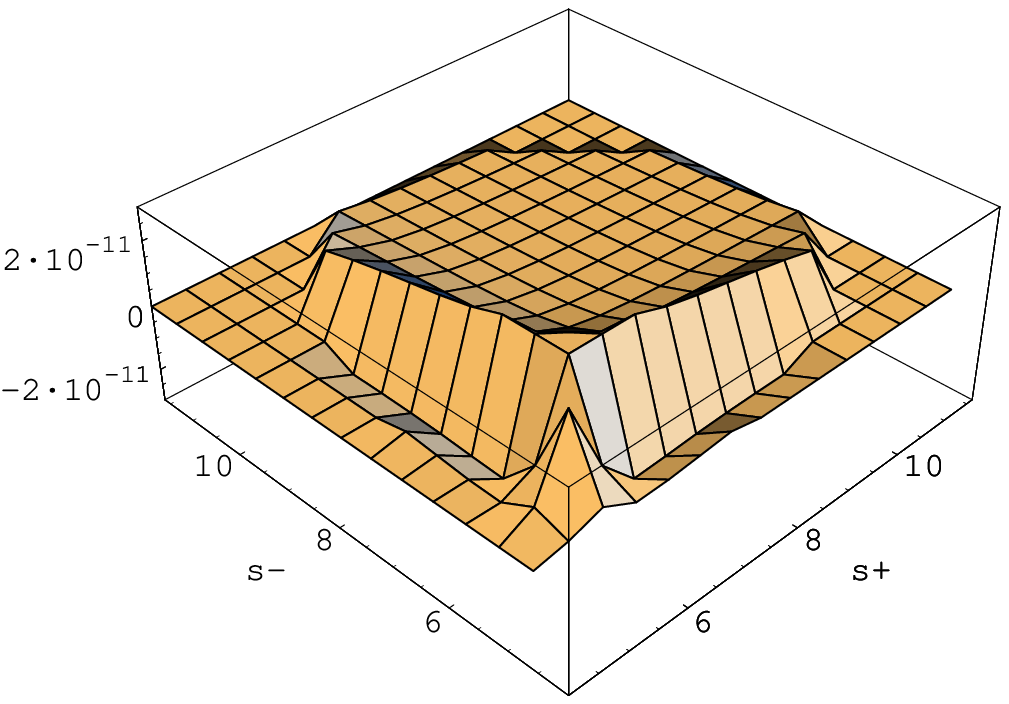,height=6cm} &
\epsfig{file=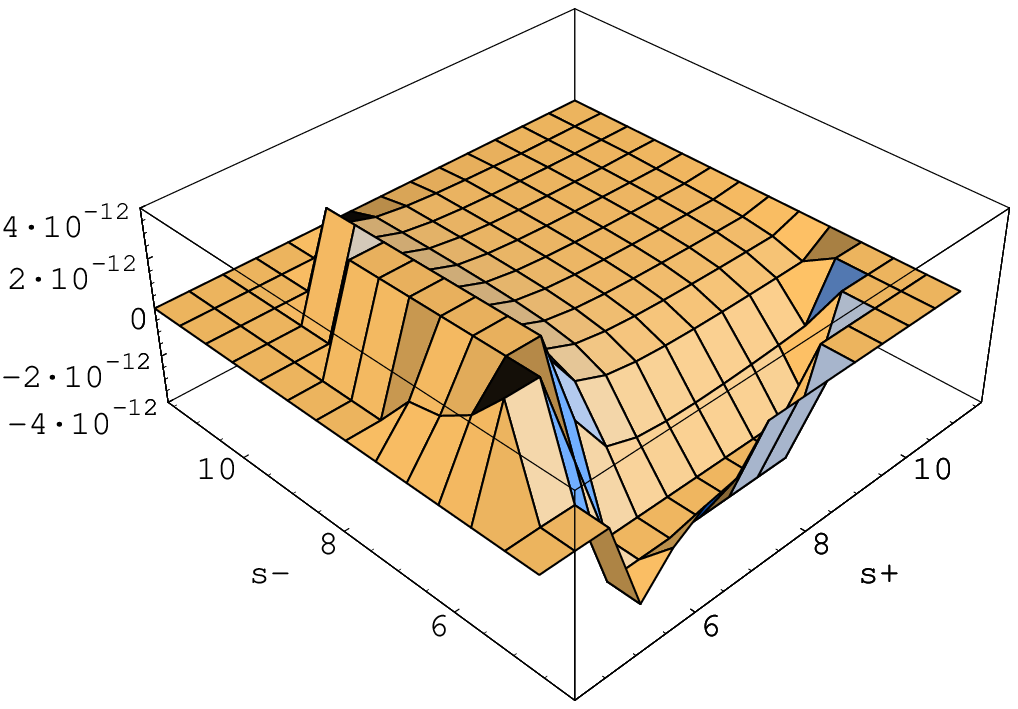,height=6cm} 
\end{tabular}
\label{f:fig3}
\caption{The functions $G_0$ (up-left), $G_c$ (up-right), 
$Re \tilde G_s$ (down-left) and $Im \tilde G_s$ (down-right) in 
Eq.(\ref{eq:gis}) using the parameterization of the amplitudes in Section 
\ref{s:efflag}.}
\end{center}
\end{figure}

We observe from both figs. 2 and 3 that the signal arising for the $D^*$ and 
$D_0$ poles cannot be  easily disentangled; concerning
the contribution of the $2^+$ pole, the $D^{*+}_2$, 
we find that it is numerically small with respect 
to the other resonances, due to the small value of the form factor
$\tau_{3/2}$ as compared to the constituent quark model value
 used in Ref. \cite{charles}.

For an assessment of the relative size of the various 
contributions to the time-dependent probabilities
 (\ref{eq:at}) and (\ref{eq:at1}), 
we integrate the functions
$G_0$, $G_c$, and $Re \tilde G_s$ and $Im \tilde G_s$ over the region
bounded by Eqs.(\ref{dalp}),(\ref{dalm}), with $s_+ \ge s_-$,
corresponding to one half of the Dalitz plot. We find, 
using the functions in fig. 3:
$\int G_0(s_+,s_-) d s_+ ds_- \simeq 7 \times 10^{-10}$ GeV$^4$,
$\int G_c(s_+,s_-) d s_+ ds_- \simeq  -7 \times 10^{-11}$  GeV$^4$,
$\int Re \tilde G_s(s_+,s_-) d s_+ ds_- \simeq  2 \times 10^{-10}$  GeV$^4$,
$\int Im \tilde G_s(s_+,s_-) d s_+ ds_- \simeq -8 \times 10^{-11}$  GeV$^4$.
These numbers may represent an indication on the sensitivity
(hence, on the required statistics) 
of a combined time-dependent and Dalitz plot analysis of the events to
the terms  
$\cos (\Delta m t)$ and
$\sin (\Delta m t)$, as well as to $\sin (2 \beta)$ and
$\cos (2 \beta)$. In particular, they suggest that the contribution 
of $\cos (2 \beta)$ 
to the time-dependent CP asymmetry may be sufficiently large to be identified.
\newpage
\section{The decay $B \to D^+ D^- K_S$}\label{s:kaon}
The kinematics of this process is quite similar
to the case of the Cabibbo suppressed $B \to D^+ D^- \pi^0$ transition 
discussed above. Apart from the change 
$m_\pi \to m_K$ which reduces the available  phase space, and the
replacement of the
$D^*$ resonance by the 
$D^*_s$ one, there are two important differences with respect to the case of 
the neutral pion in the final state. Indeed, as being induced 
at the quark level by the
$b \to c \bar c s$ transition, the amplitude for this channel is both 
color-allowed and Cabibbo favored by a factor $V_{cs}/V_{cd}$, so that one 
expects an enhanced rate of events (and, possibly, a better efficiency of 
the $K_S$ reconstruction as  compared to the $\pi^0$).
\footnote{In this regard, also the uncertainty due to the penguin 
contributions 
should be reduced.} Moreover, due to the flavour 
structure, beauty intermediate states (hence the amplitudes 
${\overline {\cal A}}_{2,3,9}$ corresponding to fig.1a) are absent, and the 
same 
is true for the $D^{*-}_s$ resonances in $\bar B_0$ decays 
(amplitudes ${\overline {\cal A}}_{5,7}$ in fig.1b). 
In addition, the $D^*_{s2}$
does not contribute in the factorization approximation. 
Therefore, the resonant 
structure of the amplitude turns out to be much simpler 
(although the parameters 
of the relevant Breit-Wigner forms are still to be measured). 
On the other hand, due to the significantly larger value of $m_K$, the 
uncertainty implicit in the application of the chiral symmetry approach is 
expected to be larger for $B \to D^+ D^- K_S$. As far as 
$SU(3)_F$ breaking effects are concerned,
we partially take them into account by replacing $f_\pi \to f_K$ 
($f_K=160$ MeV)
in the relevant formulae reported in the previous Sections.

As a result, we  obtain for  
${\overline B^0} \to D^+ D^- K_S$  the  width
$\Gamma ({\overline B^0} \to D^+ D^- K_S)\simeq 4 \times 10^{-15}$ GeV,
corresponding to the  branching fraction
${\cal B}({\overline B^0} \to D^+ D^- K_S)\simeq 9  \times 10^{-3}$.
This result indicates an enhancement of a factor 10 with respect to 
$B \to D^+ D^- \pi^0$, rather than 20 naively represented by 
$|V_{cs}/V_{cd}|^2$, and this is
a consequence of the larger value of $m_K$ reducing the
phase-space, and of
the smaller number of intermediate resonances active in this case. 

The functions relevant for the time-dependent  processes 
Eqs. (\ref{eq:at}) and (\ref{eq:at1})
are depicted in 
fig.4 which shows the expected simpler Dalitz-plot structure, as far as the 
$s_+$ and $s_-$ dependence is concerned, with respect to the process 
(\ref{process}).

The corresponding 
integrals over half of the Dalitz plot of such functions
turn out to be:
$\int G_0(s_+,s_-) d s_+ ds_- \simeq 5 \times 10^{-9}$  GeV$^4$,
$\int G_c(s_+,s_-) d s_+ ds_- \simeq  - 2 \times 10^{-9}$  GeV$^4$,
\\ \noindent
$\int Re \tilde G_s(s_+,s_-) d s_+ ds_- \simeq  2 \times 10^{-9}$  GeV$^4$,
$\int Im \tilde G_s(s_+,s_-) d s_+ ds_- \simeq - 6 \times 10^{-10}$  GeV$^4$.

\begin{figure}[ht]
\begin{center}
\begin{tabular}{cc}
\epsfig{file=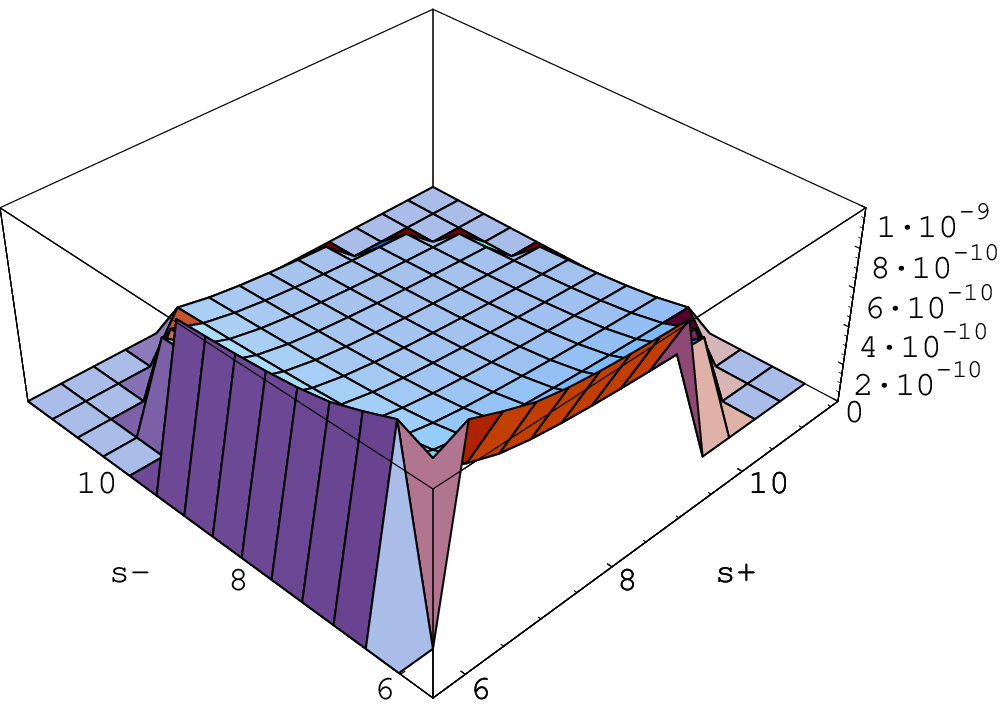,height=6cm} &
\epsfig{file=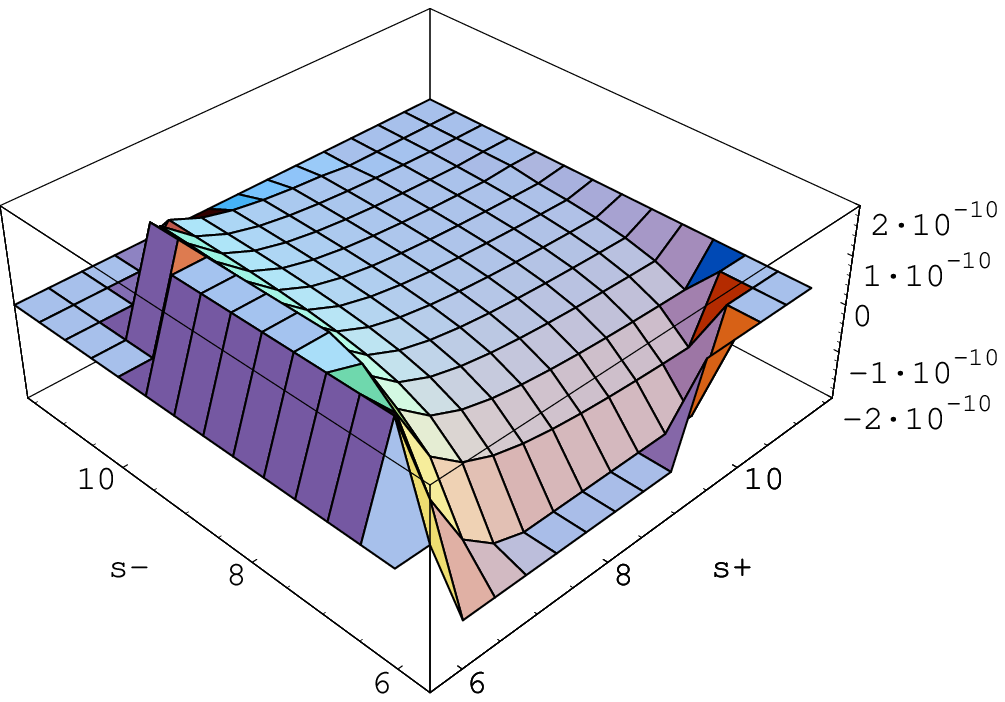,height=6cm} \\
\epsfig{file=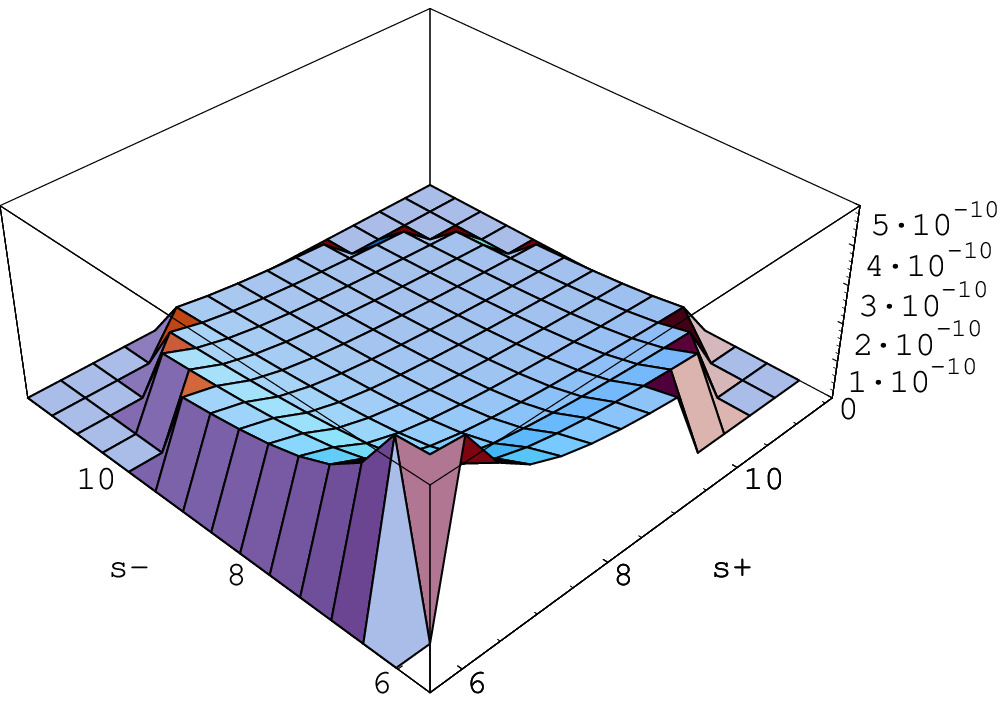,height=6cm} &
\epsfig{file=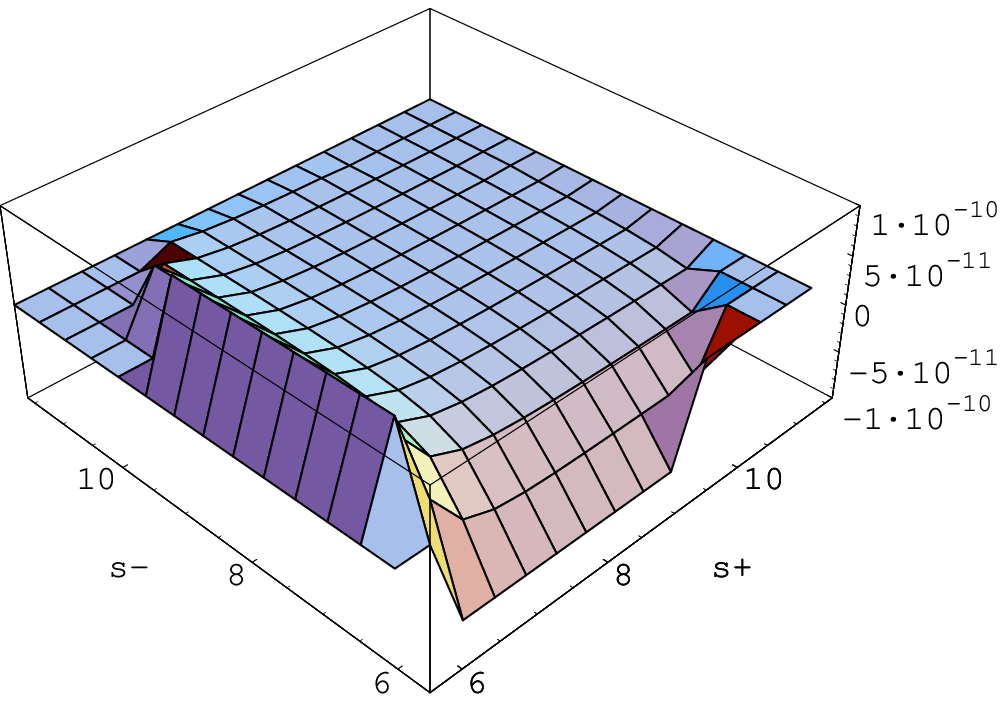,height=6cm} 
\end{tabular}
\label{f:fig4}
\caption{The functions $G_0$ (up-left), $G_c$ (up-right), 
$Re \tilde G_s$ (down-left) and $Im \tilde G_s$ (down-right) in 
Eq.(\ref{eq:gis}) for the transition $B \to D^+ D^- K_S$.}
\end{center}
\end{figure}

\section{Conclusions}\label{s:concl}

We have analyzed the three-body decays $B \to D^+ D^- \pi^0$ and
$B \to D^+ D^- K_S$ using a resonance model taking into account 
the constraints of the
chiral symmetry on the relevant transition matrix elements.
This method introduces a non-resonant, 
contact term as well as specific behaviour of the Breit-Wigner residue. 

The main conclusion of our model can be seen in figs. 2 and 3, 
which show that the coefficient of $\cos(2 \beta)$ in the time-dependent rate,
 namely $Im \tilde G_s$, 
is significantly different from zero
 over a sufficiently large portion of the Dalitz plot. 
Therefore, in principle, apart from its specific interest 
as a test of the chiral expansion in the heavy-quark theory, 
this channel might be useful to resolve the ambiguity in the 
determination of the CKM phase $\beta$ ($\beta \to {\pi \over 2} - \beta$).  
Indeed, once $\sin(2 \beta)$ is measured from, 
e.g., $B \to J/\psi K_S$, the required information should be given by a 
suitable combination according to 
Eq.(\ref{eq:at}) of the two  Dalitz plot distributions 
in the lower row of figs.2 or 3.

Qualitatively, our conclusions about the model dependence of the polar
representation and the $D^\pm$ and
 $\pi^0$ reconstruction efficiency agree with 
Ref.\cite{charles}. 
From the  numerical point of view, 
the differences with respect to \cite{charles} are mainly due to the 
inclusion of the equal-time commutator 
and the parameterization of the resonances. 
Indeed, the starting point of our calculation is the possibility of using the 
effective chiral lagrangian formalism for heavy mesons, 
offered by the smallness of the phase space available to the $\pi^0$. 
Another source of difference is the use of a smaller $\tau_{3/2}$ form factor
(as resulting from QCD sum rules) which depresses the contribution of $D_2^*$. 

Finally, we emphasize the obtained large branching ratio 
${\cal B}(B  \to D^+ D^- K_S)\simeq 9 \times 10^{-3}$ 
which, together with the simpler Dalitz plot structure and the 
better $K_S$ reconstruction efficiency,
can make this channel rather appealing for experimental analyses.

\vspace*{1cm}
\noindent {\bf Acknowledgments\\}
\noindent
(R) would like to acknowledge the support of KFUPM . He also thanks ICTP,
Trieste, for hospitality during the summer of 1998.

\begin{thebibliography}{9}
\bibitem{deshpande} 
A discussion of such effects is reported in Ch. 5 of Ref. \cite{review}.
\end{thebibliography}
\end{document}